**Title:** Parallel ICA reveals linked patterns of structural damage and fMRI language task activation in chronic post-stroke aphasia

**Authors** Joseph C. Griffis[1], Rodolphe Nenert[2], Jane B. Allendorfer[2], Jerzy P. Szaflarski[2]

**Institutional Affiliations:** University of Alabama at Birmingham Department of Psychology[1], University of Alabama at Birmingham Department of Neurology[2]

**Corresponding Author Information:** Joseph C. Griffis (joegriff@uab.edu)
Department of Neurology and UABEC, University of Alabama at Birmingham, 312 Civitan International Research Center, 1719, 6th Avenue South, Birmingham, AL 35294-0021

**Abstract**

Structural and functional MRI studies of patients with post-stroke language deficits have contributed substantially to our understanding of how cognitive-behavioral impairments relate to the location of structural damage and to the activation of surviving brain regions during language processing, respectively. However, very little is known about how inter-patient variability in language task activation relates to variability in the structures affected by stroke. Here, we used parallel independent component analysis (pICA) to characterize links between patterns of structural damage and patterns of functional MRI activation during semantic decisions using a large (N=43) sample of patients with chronic post-stroke aphasia. The pICA analysis revealed a significant association between a lesion component featuring damage to left posterior temporo-parietal cortex and the underlying deep white matter and an fMRI component featuring (1) heightened activation in a primarily right hemispheric network of frontal, temporal, and parietal regions, and (2) reduced activation bilateral areas associated with the canonical semantic network activated by healthy controls. Stronger loading parameters on both the lesion and fMRI activation components were associated with poorer language test performance. Post-hoc fiber tracking suggests that lesions affecting the left posterior temporo-parietal cortex and deep white matter may lead to the simultaneous disruption of multiple long-range structural pathways connecting distal language areas. Together, our results suggest that damage to the left posterior temporo-parietal cortex and underlying white matter may (1) impede the language task-driven recruitment of canonical left hemispheric language and other areas (e.g. the right anterior temporal lobe and default mode regions) that likely support residual language function after stroke, and (2) lead to the compensatory recruitment of right hemispheric fronto-temporo-parietal networks for tasks requiring semantic processing.

# 1. Introduction

Language impairments, commonly referred to as aphasias, are common in survivors of left hemispheric stroke. While early research relating post-stroke language impairments to the underlying neuroanatomy was limited to qualitative post-mortem examinations of individual patients (Berker et al. 1986), the development of neuroimaging analysis methods such as voxel-based lesion symptom mapping (VLSM) (Bates et al. 2003) has enabled quantitative *in vivo* characterization of lesion-deficit relationships. Similarly, the advent of functional neuroimaging techniques such as positron emission tomography (PET) and blood oxygen level-dependent (BOLD) functional MRI has enabled researchers to study how language recovery after stroke relates to language task-evoked neural activity in surviving brain regions, elucidating the neurobiological processes that support post-stroke language function (Crinion and Price 2005; Saur et al. 2006; Fridriksson et al. 2010, 2012; Szaflarski et al. 2013). However, studies focusing on characterizing relationships between language task-evoked activity in surviving brain regions and the location of structural damage are scarce.

Several studies on lesion-activation relationships have focused on identifying differences in activation between groups of patients with vs. without damage to specific anatomical regions of interest (Heiss et al. 1999; Blank et al. 2003; Turkeltaub et al. 2011). Of these, two compared brain activity between groups of patients with vs. without damage to the left inferior frontal gyrus (IFG), and their results suggest that patients with damage to the left IFG show increased activity in the contralateral right IFG during language task performance (Blank et al., 2003; Turkeltaub et al., 2011). The other (Heiss et al., 1999) compared brain activity among groups of patients with lesions affecting the basal ganglia (e.g. putamen/caudate), cortex in the anterior LMCA distribution, and cortex in the posterior LMCA distribution at different stages of recovery, and found evidence suggesting that patients with left temporal damage show increased right hemispheric activity into the chronic stage of recovery. While these studies provide important insights into lesion-activation relationships, their conclusions depend on the assumption that the observed differences in activation reflect the effects of damage to the region of interest. Although this assumption is likely valid for studies using broader ROI definitions and multiple groups (e.g. Heiss et al., 1999), the fact that lesions are not typically confined to any single anatomical region complicates the interpretation of studies where groups are defined on the basis of damage to a given anatomical region (e.g. Blank et al., 2003).

Several more recent studies have attempted to overcome this limitation by using less constrained approaches for multimodal data fusion (Specht et al. 2009; Fridriksson et al. 2010; Abel et al. 2015). One of these studies utilized VLSM to identify lesion correlates of activation summary statistics (extracted from regions where activation was associated with better naming performance) in chronic patients (Fridriksson et al., 2010). While this study was likely underpowered (N=15) to enable the identification of reliable relationships between damage and such coarse measures of activation and did not identify any lesion-activation associations at corrected statistical thresholds, results presented at liberal uncorrected statistical thresholds (i.e. $p<0.05$) suggested that weaker activation in these regions was associated with damage to the left IFG (Fridriksson et al. 2010).

Another approach has been to use a multivariate data fusion method known as multimodal joint independent component analysis (jICA) to identify joint structure-activation components that are reliably different between patients with chronic post-stroke aphasia from healthy controls (Specht et al. 2009; Abel et al. 2015). While innovative in their application of data fusion techniques to the study of post-stroke aphasia, the statistical inferences drawn from the jICA approach are based on the existence of differences between groups (i.e. stroke vs. control) in the mean mixing coefficients (i.e. subject-level loadings) for the extracted independent components (ICs) (Calhoun et al. 2006, 2009; Sui et al. 2012). The application of this approach, which decomposes information common to both MRI modalities under the assumption that they share the same mixing matrix (Calhoun et al. 2006), to functional and structural MRI data obtained from a population with major structural abnormalities (i.e. widespread tissue damage/loss) might therefore be expected to result in a biased IC selection process. This is because the joint ICs are selected, in part, on the basis that they contain features that are both (1) consistently represented across stroke patients, and (2) not consistently represented across healthy subjects (or vice versa). Intuitively, this introduces the potential for IC selection to be biased by voxel lesion frequencies, because joint ICs with lesion features containing frequently lesioned voxels are likely to differ strongly in their representation between groups. Indeed, we explicitly tested for this bias in jICA of structural and functional MRI data obtained from groups of chronic stroke patients and healthy controls, and found evidence consistent with a strong lesion-frequency bias in the jICA component selection process (Supplementary Material S1), such that voxel values for the lesion and fMRI features of the maximally differentiating joint IC were monotonic functions of group-level lesion frequencies and activation magnitudes for the stroke group, respectively. Thus, the application of jICA to identify structure-function relationships in the context of chronic post-stroke aphasia is likely to be biased towards identifying joint ICs with lesion and fMRI features that are most reliably represented across stroke patients and that are reliably absent in healthy controls.

This should not be taken to imply that the susceptibility of jICA to lesion-frequency bias invalidates the results of previous studies using jICA in stroke patients, but rather to demonstrate that jICA is likely not ideal for addressing certain types of questions in chronic stroke patients. For example, the primary hypothesis motivating the study by Specht and colleagues (2009) concerned the presence of between-group differences in right superior temporal activation between patients with temporal lesions and healthy controls. Thus, the finding that patients with temporal lesions and healthy controls differed in their contributions to the joint IC with a predominantly left temporal lesion feature and predominantly right superior temporal activation feature allows for similar but more specific conclusions than would be drawn from standard activation comparisons between the two groups.

In summary, the jICA analysis utilizes information that is shared across modalities, assumes that the structural and functional MRI features are linked to their respective sources by a single mixing matrix, and selects components based on differences in group contributions (Calhoun et al. 2006, 2009; Sui et al. 2012). In contrast, parallel ICA (pICA) is an alternative method for multimodal data fusion that only assumes that patient contributions to ICs from each modality are correlated (Liu et al. 2009a, 2009b; Sui et al. 2012). Importantly, pICA does not require that ICs be selected based on group differentiation, but rather allows for IC pairs (one from each modality) to be selected based on the strength of the correlation between their loading coefficients across patients (Liu et al. 2009b; Sui et al. 2012). Whereas the jICA approach used by previous studies attempts to answer the question "what combinations of structural and functional MRI features most reliably differ between patients and controls?," the pICA approach attempts to answer the question "what structural and functional MRI features show the most reliable relationships across patients?". Thus, while both jICA and pICA are data-driven methods for the fusion of multimodal MRI data, the component selection process for pICA is less likely to be biased by properties such as voxel lesion frequency when applied to chronic stroke patients (see Supplementary Material S1).

However, despite the advantages of pICA for investigating structure-function relationships in brain-damaged populations, to our knowledge no studies have used the pICA approach to characterize structure-function relationships in chronic post-stroke aphasia. The identification of these relationships is important for understanding how variability in the regions recruited by chronic stroke patients during language task performance relates to variability in the anatomical regions affected by stroke. A better understanding of these relationships is important for understanding of the neurobiology of language recovery after stroke, and has the potential to inform the development of experimental treatments, such as neuromodulatory interventions that are intended to induce neuroplasticity by modulating regional cortical function (Shah et al. 2013). Thus, our goal was to characterize relationships between lesion location and fMRI activation during an auditory semantic decision task. To accomplish this, we applied pICA to lesion and fMRI language task data obtained from relatively large sample of chronic stroke patients with aphasia. We note that pICA is a data-driven technique, and so the prior specification of hypotheses regarding cross-modal relationships is not necessary. Nonetheless, based on the literature discussed above (Heiss et al., 1999; Blank et al., 2003; Specht et al., 2009; Turkeltaub et al., 2011), we expected that damage to the left IFG and/or left posterior temporal cortex might be associated with increased activation in right fronto-temporal areas and/or reduced activation in canonical language networks.

## 2. Methods

*2.1 Participants*

Study procedures received approval from the Institutional Review Boards of the participating institutions. Declaration of Helsinki ethics principles and principles of informed consent were followed during all studies procedures. The current study utilized MRI and language data collected from 43 patients with chronic (> 1 year) post-stroke aphasia. All patients experienced a single left hemispheric stroke resulting in a diagnosis of aphasia at least 1 year prior to participation. Patients with right hemispheric stroke were not included in the study. All participants were screened to exclude individuals that had diagnoses of degenerative/metabolic disorders, had severe depression or other psychiatric disorders, were pregnant, were not fluent in English, or had any contraindication to MRI/fMRI.

The participants consisted of 43 patients (25 male) with a mean age of 53 (SD=15) and a mean pre-stroke Edinburgh Handedness Inventory (EHI) score of 0.85 (SD=0.43; note: EHI ranges from completely left-handed [-1] to completely right-handed [+1]) (Oldfield 1971). The mean time since stroke was 3.4 years (SD = 3.38). A detailed characterization of the patient demographics is provided in Supplementary Table 1. In addition, fMRI data from 43 age, handedness, and sex-matched controls were also analyzed to provide a reference of "typical" activation patterns during semantic decisions. This dataset is fully characterized elsewhere (Griffis et al. 2016a – Unpublished Pre-print), and additional characterization is provided in the Supplementary Material (Supplementary Table 2).

2.2 Neuroimaging data collection

MRI Data collected at the University of Alabama at Birmingham using a 3T head-only Siemens Magnetom Allegra scanner consisted of a 3D high-resolution T1-weighted anatomical scan (TR/TE = 2.3 s/2.17 ms, FOV = 25.6×25.6×19.2 cm, matrix = 256x256, flip angle = 9 degrees, slice thickness = 1mm), and two T2*-weighted gradient-echo EPI pulse functional scans (TR/TE = 2.0 s/38.0 ms, FOV = 24.0x13.6x24.0, matrix = 64x64, flip angle = 70 degrees, slice thickness = 4 mm, 165 volumes per scan). MRI data collected at the Cincinnati Children's Hospital Medical Center using a 3T research-dedicated Philips MRI scanner consisted of a 3D high-resolution T1-weighted anatomical scan (TR/TE = 8.1 s/2.17 ms, FOV = 25.0×21.0×18.0 cm, matrix = 252x211, flip angle =

8 degrees, slice thickness = 1mm) and two T2*-weighted gradient-echo EPI pulse sequence functional scans (TR/TE = 2.0 s/38.0 ms, FOV = 24.0x13.6x24.0, matrix = 64x64, flip angle = 70 degrees, slice thickness = 4 mm, 165 volumes per scan).

*2.3 MRI data processing*

All MRI data were processed using Statistical Parametric Mapping (SPM) (Friston et al. 1995) version 12 running in MATLAB r2014b (The MathWorks, Natick MA, USA). For each patient, the T1-weighted MRI scan was segmented into tissue probability maps and normalized to MNI template space using the unified normalization procedure implemented in SPM12.

Lesion probability maps were created in MNI template space using a voxel-wise naïve Bayes lesion classification algorithm implemented in the *lesion_gnb* toolbox for SPM12 (Griffis et al. 2016b). While we have shown that automated classification using this method is of comparable quality to manual lesion delineation (Griffis et al. 2016b), we chose to manually threshold the posterior probability maps in order to ensure that the resulting lesion masks precisely reflected each patient's lesion. The final lesion masks were resampled to 2 mm isotropic resolution. Figure 1A shows lesion frequencies across all 43 patients. The thresholded lesion masks were used as structural MRI inputs to the parallel ICA algorithm.

We note that our approach of using direct lesion information for the sMRI feature differs from previous studies using similar multimodal data fusion methods (i.e. JICA) in chronic stroke patients (Specht et al., 2009; Abel et al.,2015), as these studies used cerebrospinal fluid (CSF) tissue probability maps (TPMs) as proxies for lesion information. While CSF TPMs obtained from probabilistic tissue segmentation do provide relevant information about the likelihood that the tissue represented by a given voxel is lesioned (Seghier et al. 2008; Wilke et al. 2011; Guo et al. 2015; Griffis et al. 2016b), raw TPMs are sub-optimal predictors of lesion status for voxels outside of the lesion core (Griffis et al. 2016b -- see Supplementary Material). This is in part because voxels corresponding to tissues affected by gliosis and/or demyelination are often misclassified as intact tissue by automated segmentation algorithms, as their signal intensities are often similar to those observed in undamaged tissue (Mehta et al. 2003;

Seghier et al. 2008; Griffis et al. 2016b). Indeed, we explicitly compared CSF maps obtained from chronic patients to probabilistic lesion segmentations based on ground truth lesion masks, and found that CSF TPMs show relatively poor spatial similarity to the ground truth lesion masks (see Supplementary Material S2). Further, the JICA approach still demonstrated the biases described in the Introduction when CSF TPMs were used. Thus, we suggest that the use of CSF probability estimates as proxies for lesion information should be avoided by future studies in order to avoid bias and/or artifacts resulting from incomplete and/or incorrect lesion information.

Functional MRI data were pre-processed according to a standard pre-processing pipeline that consists of slice-time correction, realignment/reslicing, co-registration of the fMRI data to the anatomical image, tissue segmentation using tissue priors that are optimized for lesioned brains (Seghier et al. 2008; Ripollés et al. 2012), normalization of the anatomical scan to MNI space, normalization of the functional scan to MNI space using the transformation obtained from the normalization of the anatomical scan, and smoothing the normalized functional data with an 8mm full-width half maximum Gaussian kernel. Additionally, to reduce artifacts due to movement, the functional data were motion-corrected by replacing image volumes with >0.5mm motion with an interpolated volume from adjacent timepoints (Mazaika et al. 2005).

*2.4 Language measures*

Prior to undergoing MRI scanning, participants were administered a set of language tests that included the Boston Naming Test (BNT) (Kaplan et al. 2001), Semantic Fluency Test (SFT) (Kozora and Cullum 1995), and Controlled Oral Word Association Test (COWAT) (Lezak et al. 1995). The BNT, a picture naming test, utilizes black and white line drawings that correspond to both animate and inanimate objects. The SFT and COWAT are both fluency measures that involve generating words in response to a prompt. The SFT utilizes prompts based on semantic categories (animals, fruits/vegetables, and things that are hot), while the COWAT utilizes prompts based on letter categories (C, F, and L).

An auditory semantic decision task was administered while patients were in the scanner, providing a measure of auditory semantic comprehension (Binder et al. 1997;

Eaton et al. 2008). This task robustly activates canonical language networks (Binder et al. 1997), and is used to evoke activity related to language processing in stroke patients (Eaton et al. 2008; Szaflarski et al. 2008, 2011). During each scan, participants completed alternating blocks of semantic decision and tone decision conditions. There were five semantic decision blocks per scan, and six tone decision blocks. Each block lasted 30 seconds. During the semantic decision blocks, eight spoken English nouns designating different animals were presented, and participants decided if the animals met the criteria: "native to the United States" and "commonly used by humans". During the tone condition blocks, eight brief sequences of four to seven 500- and 750-Hz tones were presented, and participants decided if each sequence contained two 750-Hz tones. In both conditions, participants responded by pressing a button with their non-dominant hand. Each fMRI scan lasted 7 minutes and 15 seconds (165 TRs). Before performing the in-scanner task, participants confirmed their understanding of the task by completing a test run that consisted of 5 trials from each condition. Performance data were not collected for 4 patients due to hardware problems. These patients were therefore excluded from analyses involving in-scanner performance.

Patient scores on the SFT and COWAT were highly correlated ($r=0.92$). Thus, a combined fluency measure was defined as the average of the scores on each test. The combined fluency scores showed positive but less extreme correlations with naming scores ($r=0.76$) and semantic decision scores ($r=0.53$). Naming scores showed a positive correlation with semantic decision task scores ($r=0.43$). Language test scores are shown in Figure 1B. A detailed characterization of the patient test scores is provided in Supplementary Table 1.

*2.5 First-level general linear models*

Subject-level general linear models (GLM) were fit to the processed fMRI data (Friston et al. 1995). Semantic decision blocks were modeled as boxcar regressors convolved with a canonical hemodynamic response function (HRF). Temporal derivatives were included as basis functions to account for variability in the time-to-peak of the HRF (Meinzer et al. 2013). It is worth noting that because we used a blocked task design, we were not able to separately model correct vs. incorrect trials. Contrast estimate maps quantifying the difference in HRF magnitude between the semantic decision and tone decision conditions were then created for each subject. Contrast estimate maps were used as the fMRI inputs into the parallel ICA algorithm.

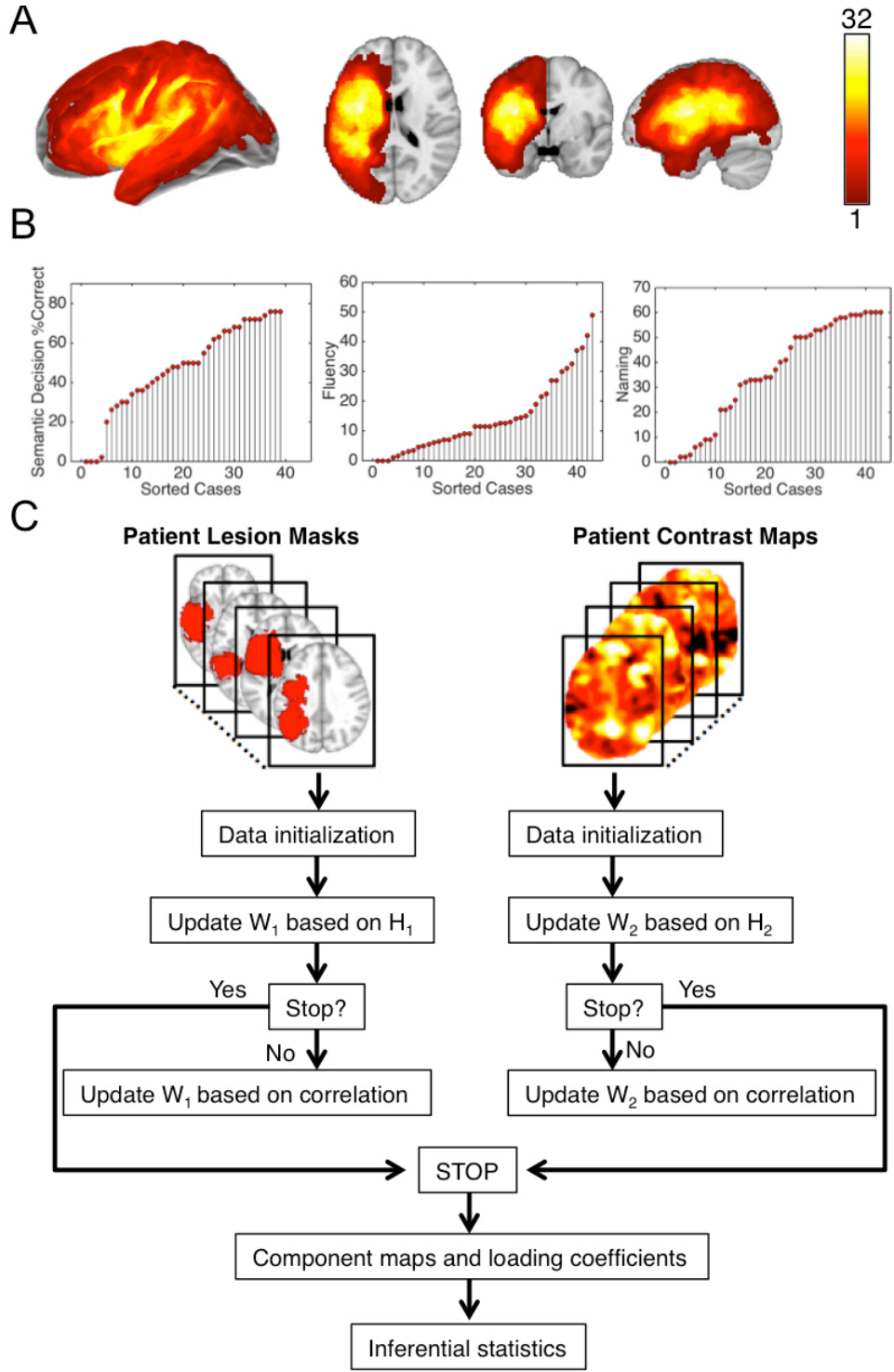

**Figure 1. Data characterization and analysis summary. A.** Lesion frequencies for all 43 patients are shown. Colorbar values indicate the number of patients with lesions at each voxel. **B.** Language test performance scores are shown (sorted from lowest to highest) for in-scanner and out of scanner language tests. Each point represents a single patient. **C.** A schematic diagram is provided to illustrate the parallel ICA procedure. ICA is performed separately on the lesion masks and contrast maps, and information about correlations between loadings on components from each modality is used to iteratively update the de-mixing matrix. Statistical inference is performed based on tests of the loading coefficients on components from each modality.

*2.6 Parallel ICA*

Parallel ICA was performed on the lesion and fMRI data using the Fusion ICA toolbox for SPM (Calhoun et al. 2006; Liu et al. 2009a). Parallel ICA is a multimodal data fusion technique that simultaneously performs ICA on each modality (Liu et al. 2009a). In addition to maximizing statistical independence among the extracted components from each modality, the correlation between loading coefficients (i.e. an index quantifying how much a given component is expressed in the data obtained from each subject) for components from each modality is used to iteratively update the de-mixing matrix, resulting in pairs of components from each modality with correlated loading coefficients across patients (Liu et al. 2009a; Sui et al. 2012). A schematic of the pICA procedure is shown in Figure 1C. Lesion data were masked to include only voxels that were lesioned in at least one patient (i.e. within the map shown in Figure 1A), and fMRI data were masked to include only in-brain voxels (Calhoun et al. 2006). Dimensionality estimation was accomplished using the minimum description length (MDL) utility implemented in the Fusion ICA toolbox (Calhoun et al. 2006; Li et al. 2007), and the number of components to be estimated was derived to be five for each modality. Five ICs were then estimated for each modality by averaging over 15 ICAs using the "Average ICA" option in the Fusion ICA toolbox with default settings (Meier et al. 2012). The resulting component pairs were then selected based on the strength of the correlation between loading coefficients (Liu et al. 2009a). Bonferroni correction was used to adjust the significance values for the cross-modal correlations by correcting for all possible cross-modal component pairings (Pearlson et al. 2015). Thus, the significance threshold was set to 0.05/25 = 0.002 to control the family-wise error rate (FWE) for the component selection process at 0.05. Spatial maps for the selected ICs were z-scaled for visualization (Meier et al., 2012).

While our sample is relatively small compared to those of several other studies that have utilized pICA to investigate structure-function relationships in clinical populations, we do note that our sample is one of the largest to date in the functional neuroimaging of post-stroke aphasia literature (see Supplementary Table 1 in Griffis et al. 2016a – Unpublished Pre-print). Nonetheless, there is the potential that our results could be disproportionately affected by the inclusion/exclusion of individual data points. Therefore, leave-one-out cross validation was performed using the tool included in the Fusion ICA toolbox in order to asses the reliability of the results when individual patients were excluded from the analysis (Liu et al. 2009b). To assess spatial similarity of the connected components obtained from the leave-one-out procedure, both the original connected component pair and each pair of connected lesion and fMRI obtained from the leave-one out procedure were thresholded at $|z|>1.96$, and converted to binary masks. The spatial similarity between each of the binary masks obtained from each of the original connected components and the binary masks obtained from each of the connected components resulting from each run of the leave-one-out procedure was then quantified using the Dice similarity coefficient (DSC), which is calculated as twice the number of voxels contained in the intersection of two images divided by the total number of voxels contained in the union of the images (Dice 1945). This metric is commonly used to assess the spatial similarity between images, and values of 0.6 and higher are typically considered indicative of achieving "good" similarity between two images (Zou et al. 2004; Seghier et al. 2008; Wilke et al. 2011; Griffis et al. 2016b).

*2.7 Imaging-behavior relationships*

Relationships between patient language test scores and patient loadings on the extracted fMRI and lesion ICs were explored with robust multiple linear regressions using iteratively re-weighted least squares (using the default bi-square weighting function implemented in the MATLAB Statistics Toolbox) to reduce the effect of outlier data points (Poldrack 2012). In addition to using robust regression to reduce the impact of outliers, we also ran additional analyses using outlier exclusion criteria (cases falling >2 SD from the regression line) that were employed by a previous study on a similar topic (Fridriksson et al. 2009).

Because the components that would be derived from the lesion and fMRI data could not be known prior to performing the pICA analysis, *a priori* predictions about their relationships to behavioral measures could not be made. Thus, while these analyses should be considered exploratory in this regard, they are important for understanding how patient contributions to each of the identified lesion and fMRI patterns relate to language abilities.

For the lesion data, three robust regression models were fit (one for each language measure) using the standardized component loadings on all 5 lesion ICs as predictors and the standardized language scores as the outcome variables. For the fMRI data, three robust regression models were fit (one for each language measure) using the standardized component loadings on all 5 fMRI ICs as predictors and the standardized language scores as the outcome variables. Models and parameter estimates are presented with p-values and the expected false discovery rate (FDR) (Benjamini and Hochberg 1995; Genovese et al. 2002). FDR values at a given threshold express the estimated proportion of discoveries at that threshold that are expected to be false. Of the 36 tests performed in this analysis, 14 tests were significant at an FDR threshold of 0.1, suggesting that no more than 1-2 discoveries at this threshold are expected to be false positives. We considered this to be acceptable, given the data-driven nature of the analyses presented. Lastly, while we used robust regression models to reduce the influence of outliers on our results, each model with outlier cases meeting the criteria (i.e. cases greater than 2 SD from the regression line) employed by Fridriksson and colleagues (2009) was re-ran with these cases excluded to further confirm that our results were not being driven by outlier cases.

*2.8 Additional analyses*

Additional analyses were performed to aid in the interpretation of the results of the pICA and imaging-behavior analyses. First, additional fMRI semantic decision task data were obtained from a group of 43 age, handedness, and sex-matched controls as mentioned above. This dataset is fully described in a separate report (Griffis et al. 2016a – Unpublished Pre-print), and are also described in Supplementary Material 1. The control fMRI data were analyzed using a standard mass univariate dependent samples t-contrast as implemented in SPM12; the resulting group-level activation map that was thresholded at a voxel-wise $p<0.01$, and cluster-corrected at $p<0.05$ (k = 99). This was used as a reference for what constitutes a "typical" activation pattern during the SDTD task.

Secondly, deterministic tractography was performed to characterize the white matter connections likely to be affected by damage to regions with strong ($|Z|>3.09$) contributions to lesion ICs identified by the pICA analysis.
We used a freely available tractography atlas (WU-Minn HCP Consortium; HCP-842 atlas - http://dsi-studio.labsolver.org/download-images/hcp-842-template) constructed using diffusion MRI data from the Human Connectome Project (2015 Q4, 900 subject release, 842 subjects included). Data were accessed under the WU-Minn HCP open access agreement, and were originally acquired using a multi-shell diffusion scheme (b-values: 1000, 2000, and 3000 s/mm2; diffusion sampling directions: 90, 90, and 90; in-plane resolution: 1.25mm). Q-space diffeomorphic reconstruction (QSDR) as implemented in DSI_Studio (Yeh and Tseng 2011) was used to reconstruct the data in MNI template space and obtain the spin distribution function (Yeh et al. 2010) (diffusion length sampling ratio: 1.25; output resolution: 2mm). Deterministic fiber tracking (Yeh et al. 2013) seeded the whole brain to calculate 100,000 tracts (constrained to terminate within the 20% thresholded grey matter probability map included with SPM12). We used the default tracking parameters implemented in DSI_studio (angular threshold: 60 degrees; step size: 1 mm; quantitative anisotropy threshold determined automatically by DSI Studio to be 0.24; tracks with length less than 30 mm were discarded). The resulting tracts were filtered to leave only tracts that passed through voxels with dominant contributions ($|Z|>3.09$) to the lesion ICs of interest. The filtered tracts were manually separated and labeled according to previous reports (Catani et al. 2002; Catani and Mesulam 2008; Catani and Thiebaut de Schotten 2008; Hua et al. 2008; Turken and Dronkers 2011).

# 3. Results

## 3.1 Parallel ICA correlation results

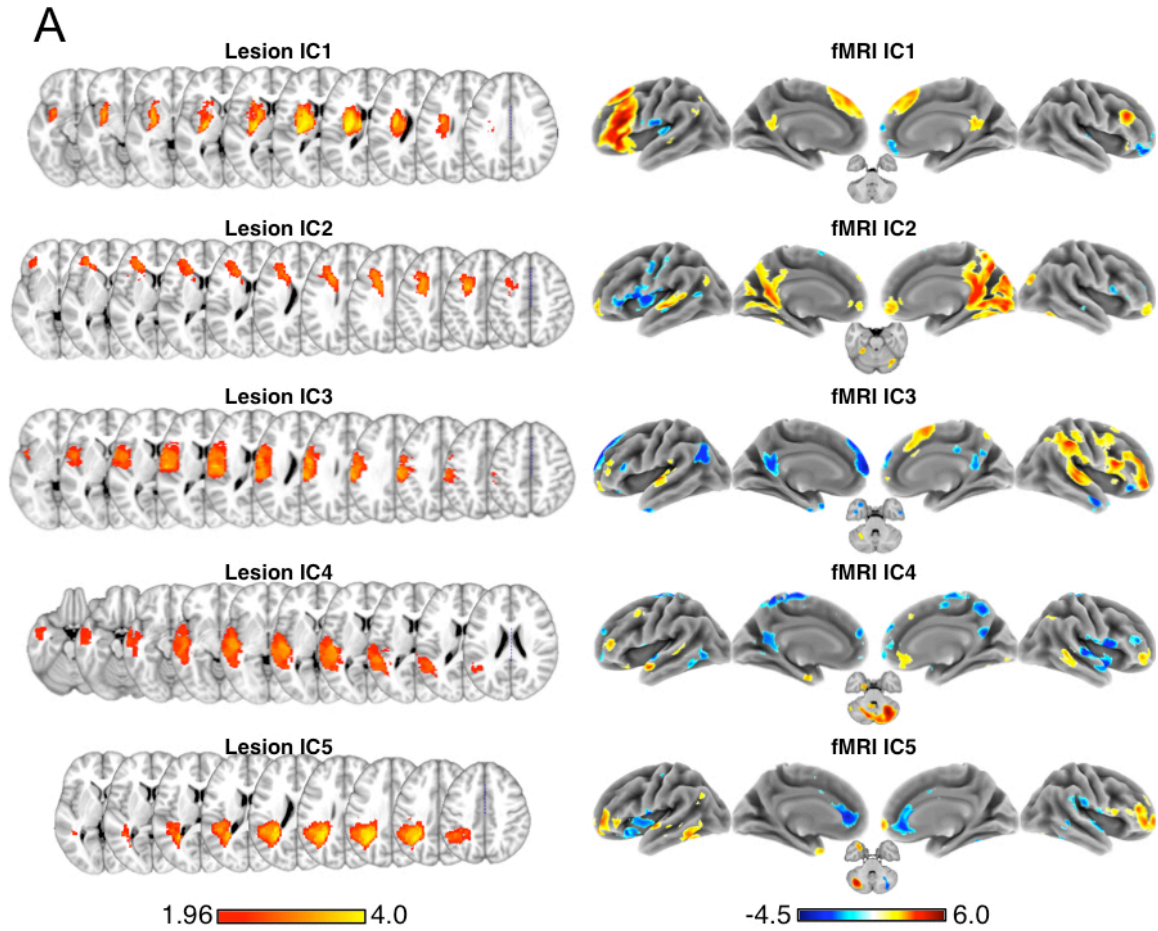

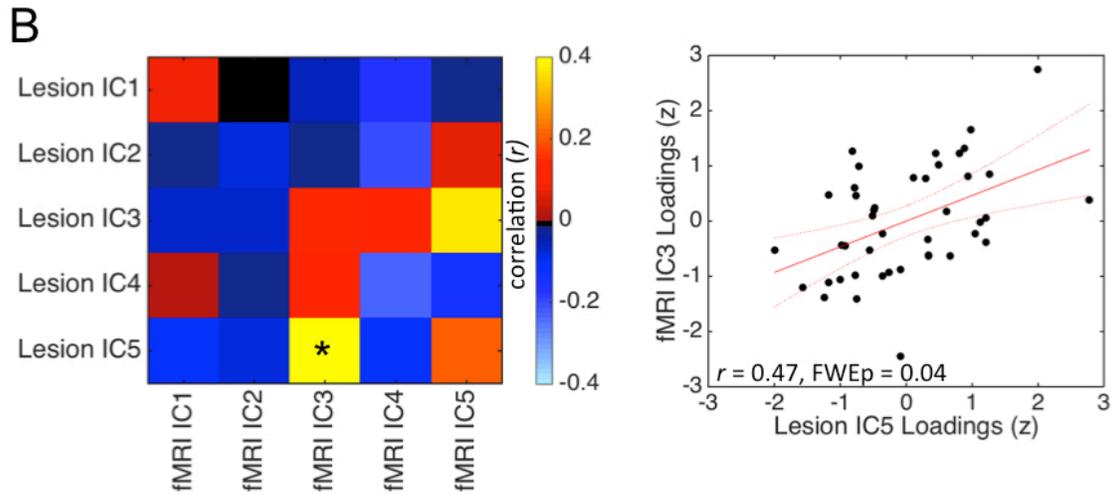

**Figure 2. Parallel ICA Results. A.** The thresholded (|z|>1.96) lesion (left) and fMRI (right) ICs obtained from the pICA analysis are shown. Colorbar values indicate z-scaled voxel contributions. **B.** Left – correlations between each lesion and fMRI IC are shown in a correlation matrix, where colorbar values indicate linear (Pearson) correlation values and significant (FWEp<0.05) inter-modal correlations are marked with an asterisk. Right – a scatterplot illustrating the relationship between z-scaled patient component loadings on significantly linked components lesion IC3 and fMRI IC5.

Results from the pICA analysis are shown in Figure 2A. The pICA analysis revealed a single pair of ICs (Lesion IC5 and fMRI IC3) that had significantly correlated loading coefficients across patients ($r$=0.47, $p$=0.04, corrected). A scatterplot illustrating the relationship between patient loading coefficients on lesion IC5 and fMRI IC3 is shown in Figure 2B. Partial correlation analyses confirmed that component loadings remained significantly correlated even when lesion volume effects (partial $r$=0.40, $p$=0.009) and scanner site (partial $r$=0.45, $p$=0.003) were partialled out.

The leave-one-out cross-validation revealed an average between-modality correlation of 0.46 (SD=0.08) for the most connected component pair. The mean DSC between the original and leave-one-out lesion components of these pairs was 0.75 (SD=0.29), and the mean DSC between the original and leave-one-out fMRI components from these pairs was 0.80 (SD=0.14). Thus, the leave-one-out procedure produced similarly correlated component pairs that had high spatial similarity to those obtained from the original pICA on the full dataset, attesting to the stability of our results.

The corresponding spatial maps for each IC, which consist of z-scaled loading parameters at each voxel, were thresholded to retain voxels with |z| > 1.96 (i.e. $p$<0.05) in order to emphasize the dominant voxel contributions to each component (Figure 2A). Lesion IC5 featured dominant contributions from posterior temporo-parietal cortex and the underlying white matter (Figure 2A). FMRI IC3 featured dominant negative contributions from the left inferior frontal gyrus pars triangularis (IFGptr), left angular gyrus, bilateral superior frontal gyri (SFG), bilateral anterior temporal lobes (ATL), bilateral posterior cingulate cortex (PCC) and precuneus, right middle cingulate cortex, and the right inferior frontal gyrus pars orbitalis (IFGporb) (Figure 2A; Table 1). FMRI

IC3 also featured dominant positive contributions from a set of predominantly right hemispheric frontal, temporal, and parietal cortices that included the right inferior parietal lobule (IPL), right superior temporal gyrus (STG), right IFGptr, right supplementary motor area (SMA), and left cerebellum (Figure 2A, Table 1). Cluster and peak statistics for the pair of significantly correlated ICs are shown in Table 1.

**Table 1. Cluster and peak statistics for linked components identified by parallel ICA.**

| fMRI IC 3 | | | | | |
|---|---|---|---|---|---|
| Region | Extent | z-value | x | y | z |
| R Inferior Parietal Lobule | 3427 | 6.16 | 46 | -42 | 52 |
| R Superior Temporal Gyrus | 3427 | 3.83 | 62 | -28 | 4 |
| R IFG (p. Triangularis) | 4910 | 4.48 | 52 | 20 | 24 |
| R Middle Orbital Gyrus | 4910 | 4.00 | 36 | 52 | -6 |
| R Posterior-Medial Frontal | 4910 | 3.73 | 4 | 12 | 52 |
| L Superior Temporal Gyrus | 301 | 3.92 | -56 | -14 | 2 |
| L Cerebelum (VI) | 383 | 3.24 | -30 | -56 | -28 |
| L Superior Temporal Gyrus | 198 | 2.85 | -50 | -30 | 10 |
| R Thalamus | 88 | 2.69 | 6 | -16 | 4 |
| L Middle Orbital Gyrus | 88 | 2.65 | -26 | 56 | -10 |
| R Precuneus | 33 | 2.32 | 10 | -68 | 54 |
| L Superior Medial Gyrus | 1997 | -4.28 | -4 | 58 | 12 |
| L Superior Frontal Gyrus | 1997 | -4.07 | -14 | 48 | 44 |
| Cerebellar Vermis (4/5) | 551 | -3.56 | -2 | -48 | 8 |
| R Middle Temporal Gyrus | 221 | -3.12 | 62 | -2 | -22 |
| L Angular Gyrus | 591 | -3.01 | -42 | -62 | 28 |
| L IFG (p. Triangularis) | 71 | -2.84 | -48 | 28 | 16 |
| R IFG (p. Orbitalis) | 86 | -2.78 | 48 | 32 | -14 |
| L Inferior Temporal Gyrus | 92 | -2.48 | -40 | -4 | -38 |
| R MCC | 26 | -2.41 | 2 | -28 | 34 |
| L Medial Temporal Pole | 58 | -2.28 | -32 | 12 | -34 |
| L Hippocampus | 37 | -2.24 | -28 | -14 | -10 |
| Lesion IC5 | | | | | |
| Region | Extent | z-value | x | y | z |
| Supramarginal Gyrus (Deep White Matter) | 4103 | 3.87 | -30 | -48 | 30 |
| Middle Temporal Gyrus | 4103 | 2.12 | -46 | -46 | 4 |
| Superior Temporal Gyrus (Deep White Matter) | 4103 | 1.97 | -44 | -44 | 12 |

*Note: statistics are shown for clusters with a minimum of 20 voxels.*

*3.2 Lesion-behavior relationships*

A robust regression model was fit using lesion IC loadings to predict fluency scores ($R^2=0.35$, $F_{5,37}=3.92$, p=0.005, FDR=0.036). Lesion IC2 and lesion IC5 loadings uniquely predicted fluency scores at an FDR threshold of 0.1 (Table 2; Figure 3A). Two cases met the exclusion criteria employed by Fridriksson and colleagues (2009), and a second model was fit without these cases ($R^2=0.40$, $F_{5,35}=4.62$, p=0.002). After excluding these cases, lesion IC5 loadings (ß=-0.51, t=-3.92, p=0.0003) and lesion IC2 loadings (ß=-0.37, t=-2.83, p=0.007) remained the only unique predictors of fluency scores at a per-comparison p<0.05.

A robust regression model was fit using lesion IC loadings to predict naming scores ($R^2=0.31$, $F_{5,37}=3.36$, p=0.01, FDR=0.045), Lesion IC 5 loadings uniquely predicted naming scores at an FDR threshold of 0.1 (Table 2; Figure 3A). Two cases met the exclusion criteria employed by Fridriksson and colleagues (2009), and a second model was fit without these cases ($R^2=0.43$, $F_{5,35}=5.18$, p=0.001). After excluding these cases, lesion IC5 loadings remained a unique predictor of naming scores (ß=-0.32, t=-2.17, p=0.036), although lesion IC3 loadings were also a unique predictor of naming scores (ß=-0.36, t=-2.62, p=0.01) a per-comparison p<0.05.

A robust regression model was fit using lesion IC loadings to predict AudSem scores ($R^2=0.23$, $F_{5,33}=1.93$, p=0.11, FDR=0.23). Lesion IC 5 loadings uniquely predicted AudSem scores at an FDR threshold of 0.1 (Table 2; Figure 3A). Four cases that met the exclusion criteria employed by Fridriksson and colleagues (2009), and a second model was fit without these cases ($R^2=0.40$, $F_{5,29}=3.91$, p=0.008). After excluding these cases, lesion IC5 loadings remained the only unique predictor of AudSem scores (ß=-0.43, t=-3.24, p=0.003) at a per-comparison p<0.05.

**Table 2. Robust regression results for lesion ICs (all cases included).**

| Fluency (Lesion IC predictors) | | | |
|---|---|---|---|
| Predictor | ß estimate | t-statistic | p-value (FDR) |
| Lesion IC1 | -0.13 | -0.88 | 0.39 (0.56) |
| **Lesion IC2** | **-0.44** | **-2.9** | **0.006 (0.036)** |
| Lesion IC3 | -0.06 | -0.45 | 0.65 (0.73) |
| Lesion IC4 | -0.05 | -0.41 | 0.68 (0.74) |
| **Lesion IC5** | **-0.5** | **-3.28** | **0.002 (0.024)** |

| Naming (Lesion IC predictors) | | | |
| --- | --- | --- | --- |
| Predictor | ß estimate | t-statistic | p-value (FDR) |
| Lesion IC1 | -0.1 | -0.66 | 0.51 (0.63) |
| Lesion IC2 | -0.25 | -1.5 | 0.14 (0.27) |
| Lesion IC3 | -0.25 | -1.63 | 0.11 (0.23) |
| Lesion IC4 | -0.22 | -1.51 | 0.13 (0.26) |
| **Lesion IC5** | **-0.34** | **-2.1** | **0.04 (0.096)** |
| AudSem (Lesion IC predictors) | | | |
| Predictor | ß estimate | t-statistic | p-value (FDR) |
| Lesion IC1 | 0.08 | 0.45 | 0.65 (0.73) |
| Lesion IC2 | -0.13 | -0.75 | 0.46 (0.61) |
| Lesion IC3 | 0.09 | 0.52 | 0.61 (0.73) |
| Lesion IC4 | -0.21 | -1.25 | 0.22 (0.38) |
| **Lesion IC5** | **-0.42** | **-2.33** | **0.03 (0.09)** |

*Note: Bold text indicates significant predictors at a per-comparison p<0.05 (FDR<0.1)*

*3.3 fMRI-behavior relationships*

A robust regression model was fit using fMRI IC loadings to predict fluency scores with scanner included as a covariate of no interest ($R^2=0.53$, $F_{6,37}= 6.63$, p<0.0001, FDR=0.0001). fMRI IC1 and fMRI IC3 loadings uniquely predicted fluency scores at an FDR threshold of 0.1 (Table 3; Figure 3B). Two cases met the exclusion criteria employed by Fridriksson and colleagues (2009), and a second model was fit without these cases ($R^2=0.42$, $F_{6,34}=4.08$, p=0.003). After excluding these cases, fMRI IC1 loadings (ß=0.53, t=3.60, p=0.0009) and fMRI IC3 loadings (ß=-0.33, t=-2.79, p=0.008) remained the only unique predictors of fluency scores at a per-comparison p<0.05.

A robust regression model was fit using fMRI IC loadings to predict naming scores with scanner included as a covariate of no interest ($R^2=0.29$, $F_{6,37}= 2.47$, p=0.04, FDR=0.096). fMRI IC1 and fMRI IC3 loadings uniquely predicted naming scores at an FDR threshold of 0.1 (Table 3; Figure 3B). No cases met the exclusion criteria employed by Fridriksson and colleagues (2009), and so no additional models were fit.

A robust regression model was fit using fMRI IC loadings to predict AudSem scores with scanner included as a covariate of no interest ($R^2=0.43$, $F_{6,32}= 4.05$, p=0.003, FDR=0.027). fMRI IC1 and fMRI IC2 loadings uniquely predicted AudSem scores at an FDR threshold of 0.1 (Table 3; Figure 3B). Two cases met the exclusion criteria employed by Fridriksson and colleagues (2009), and a second model was fit without these cases ($R^2=0.48$, $F_{6,30}=4.62$, p=0.002). After excluding these cases, fMRI IC1 loadings (ß=0.38, t=2.82, p=0.008) and fMRI IC2 loadings (ß=0.40, t=-2.99, p=0.006) remained the only unique predictors of fluency scores at a per-comparison p<0.05.

**Table 3. Robust regression results for fMRI ICs (all cases included).**

| Predictor | ß estimate | t-statistic | p-value (FDR) |
|---|---|---|---|
| Fluency (fMRI IC predictors) | | | |
| **fMRI IC1** | **0.63** | **4.96** | **<0.0001 (0.0001)** |
| fMRI IC2 | -0.02 | -0.17 | 0.86 (0.88) |
| **fMRI IC3** | **-0.33** | **-2.62** | **0.01 (0.045)** |
| fMRI IC4 | 0.16 | 1.32 | 0.19 (0.34) |
| fMRI IC5 | 0.08 | 0.68 | 0.5 (0.63) |
| Naming (fMRI IC predictors) | | | |
| **fMRI IC1** | **0.41** | **2.51** | **0.02 (0.065)** |
| fMRI IC2 | 0.19 | 1.23 | 0.23 (0.37) |
| **fMRI IC3** | **-0.34** | **-2.1** | **0.04 (0.096)** |
| fMRI IC4 | 0.19 | 1.18 | 0.25 (0.39) |
| fMRI IC5 | -0.02 | -0.15 | 0.88 (0.88) |
| AudSem (fMRI IC predictors) | | | |
| **fMRI IC1** | **0.38** | **2.5** | **0.02 (0.065)** |
| **fMRI IC2** | **0.39** | **2.6** | **0.02 (0.065)** |
| fMRI IC3 | -0.04 | -0.25 | 0.8 |
| fMRI IC4 | 0.21 | 0.9 | 0.37 |
| fMRI IC5 | 0.2 | 0.82 | 0.42 |

*Note: Bold text indicates significant predictors at a per-comparison p<0.05 (FDR<0.1)*

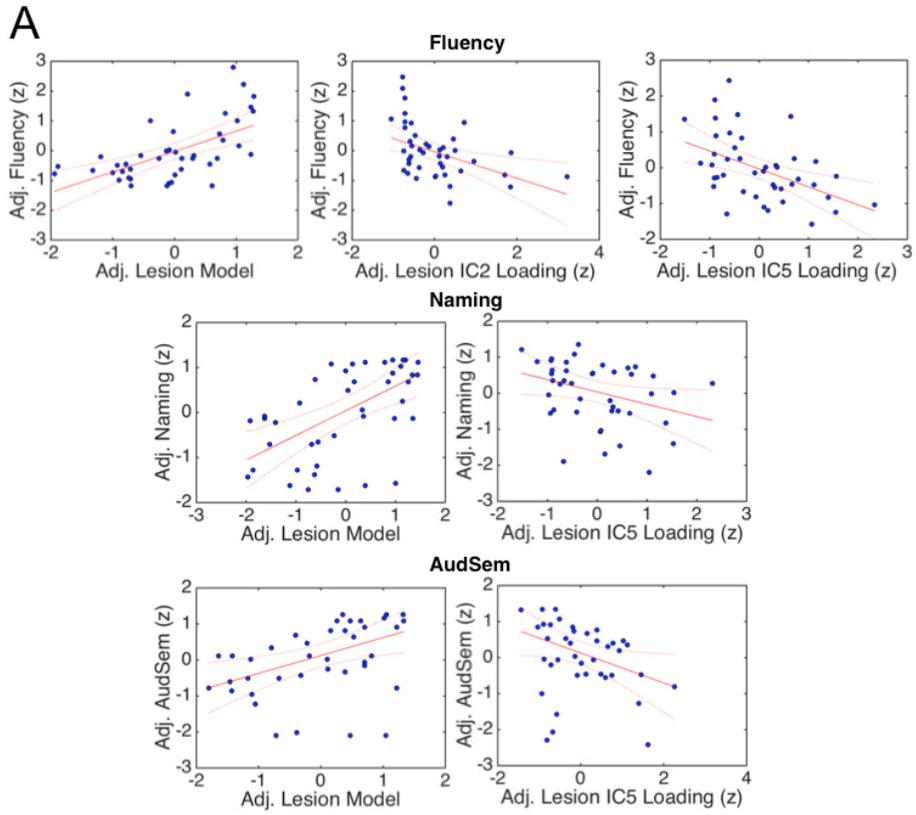
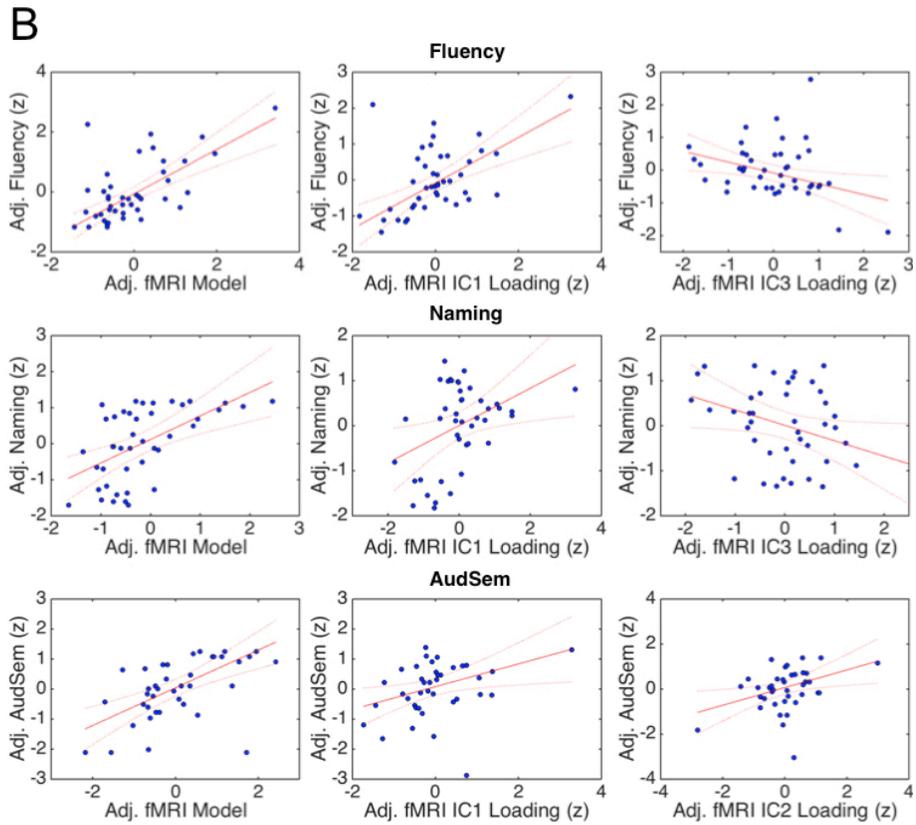

**Figure 3. Imaging-behavior relationships. A.** The fitted robust regression (using all cases) models using lesion IC loadings to predict scores on each language measure are shown on the left-most panel of each row, and added variable plots illustrating the unique effect of each significant lesion IC predictor are shown to the right. **B.** The fitted robust regression (using all cases) models using fMRI IC loadings to predict scores on each language measure are shown on the left-most panel of each row, and added variable plots illustrating the unique effect of each significant fMRI IC predictor are shown to the right.

*3.4 Qualitative comparisons of fMRI ICs 3 and 5 to typical activation patterns*

The activation map obtained from the mass univariate analysis of the control fMRI data is shown in Figure 4A, along with voxel contributions two fMRI ICs that showed opposite relationships to behavior (fMRI IC1 and fMRI IC3). Qualitatively, fMRI IC1 (which showed positive relationships to all three language measures) featured positive contributions from several regions activated by controls, and negative contributions from several regions de-activated by controls. In contrast, fMRI IC3 (which showed a positive relationship to lesion IC5, and negative relationships to fluency and naming) featured negative contributions from several regions activated by controls, and positive contributions from several regions deactivated by controls. While qualitative, this supports the interpretation that fMRI IC1 reflects activity in the typical semantic network, whereas fMRI IC3 reflects activity in an atypical network during semantic processing.

*3.5 Deterministic fiber tracking of lesion ICs 4 and 5*

Patient component loadings for only one (lesion IC5) of the two (lesion IC4 and lesion IC5) lesion components featuring dominant contributions from temporal lobe voxels showed significant relationships to fMRI activation (Figure 2) or language task performance (Table 2; Figure 3A). A close examination of the voxels contributing to each component reveals an important distinction. While lesion IC4 featured the most prominent ($|z|>3.09$) contributions from voxels in the superior temporal gyrus, these primarily came from cortical and/or superficial white matter voxels (Figure 4B). In contrast, lesion IC5 featured prominent ($|z|>3.09$) contributions from voxels in the deep temporal and parietal white matter (Figure 4B). One potential reason for the differential effects of these lesion patterns on task activation and language test performance is that they have different effects on inter-regional connections. Indeed, deterministic fiber tracking was performed as described in the Methods section, allowing for the identification of white matter tracts passing through the voxels with the most dominant ($|z|>3.09$) contributions to each IC. Figure 4B shows that while only fibers associated with a single tract (the middle longitudinal fasciculus – MdLF) were found to pass through the voxels with the most dominant contributions to lesion IC4, fibers associated with seven different long-range tracts were found to pass through the voxels with the most dominant contributions to lesion IC5. Notably, the tracts passing through the voxels with the most dominant contributions to lesion IC5 included the long segment of the arcuate fasciculus (AFls), the posterior segment of the arcuate fasciculus (AFps), the anterior segment of the arcuate fasciculus (AFas), and the inferior fronto-occipital fasciculus (IFOF), and each of these pathways connects language-relevant cortices and supports various aspects of language function (Catani and Mesulam 2008; Fridriksson et al. 2013; Forkel et al. 2014; Ivanova et al. 2016). This supports the interpretation that lesion patterns reflected by lesion IC5 may be more likely to result in the large-scale disconnection of language-relevant cortices than lesion patterns reflected by lesion IC4, although this cannot be definitely concluded from the current analyses.

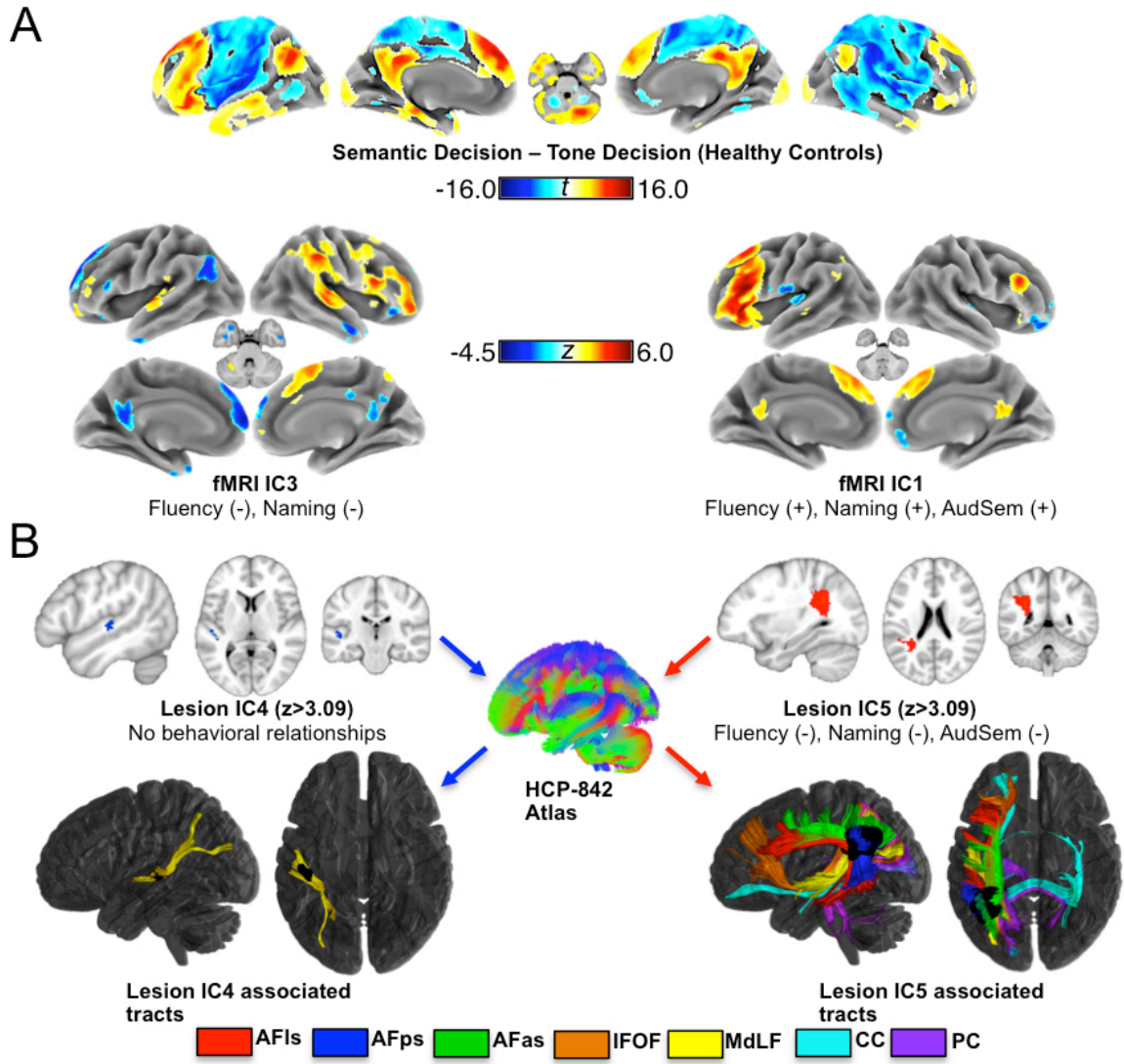

**Figure 4. Post-hoc characterization of selected fMRI and lesion components. A.** Regional activations (warm colors) and deactivations (cool colors) associated with the semantic decision task in a group of 43 age, sex, and handedness matched controls (voxel p<0.01, cluster-corrected p<0.05) are shown along with dominant (|z| > 1.96) voxel contributions to fMRI IC3 (left) and fMRI IC1 (right), which showed opposite behavioral relatisonhips (as indicated by the text under the IC label). **B.** Stringently thresholded (|Z|>3.09) maps corresponding to lesion IC4 (blue) and lesion IC5 (red) are shown to emphasize voxels with the strongest contributions to each IC. Note that dominant contributions to lesion IC4 are primarily from cortical and/or superficial white matter voxels in the superior temporal lobe, while dominant contributions to lesion IC5 are located primarily in the deep temporo-parietal white matter. Deterministic fiber tracking

using the HCP-842 atlas with ROI filters corresponding to the stringently thresholded maps for lesion ICs 4 and 5 reveals differences in their capacity for disrupting structural connectivity.

## 4. Discussion

*4.1 Lesion effects on functional activation*

Structural neuroimaging studies have enabled important advances in understanding how language deficits after stroke relate to the anatomical regions that are damaged. Similarly, functional neuroimaging studies have provided important insights into how language recovery after stroke relates to the modulation of activity in surviving brain regions during language task performance. However, very few studies have investigated how differences in language task-driven activations among patients with chronic post-stroke aphasia relate to differences in the areas damaged by stroke. Previous studies addressing this question have been constrained by arbitrary group definitions (Heiss et al. 1999; Blank et al. 2003) and have faced limitations related to small sample sizes and methodological factors (Specht et al. 2009; Fridriksson et al. 2010; Abel et al. 2015). Here, we attempted to overcome these limitations by using pICA – a recently developed technique for the simultaneous analysis of multimodal neuroimaging data – to identify linked patterns of structural damage and fMRI activation associated with auditory semantic decision task performance in a comparatively large sample of patients with chronic post-stroke aphasia. Our results indicate that lesion patterns involving damage to the left posterior temporo-parietal cortex and underlying deep white matter (lesion IC5) are associated with an activation pattern (fMRI IC3) consisting of (1) reduced activation in fronto-parietal and PCC regions that are typically recruited during semantic decisions, and (2) increased activation in a primarily right-lateralized frontal/parietal/temporal network of regions that are not typically recruited during semantic decisions. Additional analyses revealed that larger contributions to these lesion and fMRI activation patterns were associated with worse language abilities. Exploratory analyses using deterministic fiber tracking suggest that the presence of atypical fMRI activation patterns and broad language deficits in patients with lesion patterns in the area of the left posterior temporo-parietal cortex and underlying white matter resembling lesion IC5 may be due, in part, to the simultaneous disruption of multiple long-range white matter pathways that are known to support various language processes. We discuss these results in the context of the broader literature in the following sections.

We first note that the semantic decision task used in this study has been consistently shown to produce robust activation in a bilateral but predominantly left hemispheric canonical semantic network that includes the IFGptr and IFGporb, MFG, SFG, ATL, posterior inferior temporal gyrus (pITG), angular gyrus, precuneus, and posterior cingulate cortex (PCC) in healthy individuals (see Figure 4A) (Binder et al. 1997, 1999, 2008, 2009; Eaton et al. 2008; Szaflarski et al. 2008; Donnelly et al. 2011; Kim et al. 2011). In addition to co-activating during semantic decisions, the regions that form this canonical semantic network are structurally connected via pathways that include the AF, IFOF, and MdLF, and also show correlated activity fluctuations during rest (Turken and Dronkers 2011). Previous work suggests that successful language recovery after left hemispheric stroke depends, in part, on the renewed recruitment of surviving regions associated with such canonical networks during language task performance (Heiss and Thiel 2006; Saur et al. 2006; Saur and Hartwigsen 2012; Szaflarski et al. 2013). Accordingly, we observed that activation patterns resembling fMRI IC1, which featured activation in regions associated with the canonical semantic network (Figure 4A), were associated with better performance on all three language measures (Table 3; Figure 3B).

However, the pICA results indicate that left posterior temporo-parietal lesions resembling lesion IC5 were associated with a different activation pattern that consisted of (1) reduced activity in canonical semantic network regions including the left IFGptr, left mSFG, left AG, left hippocampus, bilateral ATL, and bilateral PCC/precuneus (Figure 2A; Figure4A; Table 1), and (2) increased activity in non-canonical right hemispheric regions that include the right IFGptr, right IFGpop, right pre-central gyrus, right supplementary motor area, and right IPL/SMG (Figure 2A; Figure 4A; Table 1). Importantly, loadings on this activation pattern showed negative relationships to performance on tests of both picture naming and category fluency (Table3; Figure 3B). Right hemispheric regions (particularly the right IFGpop and right SMA) have been previously found to show increased activation during auditory language task performance in acute patients that normalize with successful long-term recovery (Saur et al. 2006). Thus, it is thought that right hemispheric compensation is most pronounced immediately following the abrupt disruption of left hemispheric function by the stroke, and that the normalization of right hemispheric activation during recovery is reflective of a renewed capacity of surviving left hemispheric areas to support language processing (Saur et al., 2006; Heiss and Thiel 2006; Saur and Hartwigsen, 2012). Along these lines, it has been proposed that increases in right hemispheric activation that persist into the chronic recovery stage reflect the maintenance of early compensatory mechanisms driven by a failure to restore left hemispheric involvement in language processing (Heiss and Thiel, 2006; Saur and Hartwigsen, 2012). Under this proposal, our results suggest that lesion patterns resembling lesion IC5 (i.e. the left posterior temporo-parietal cortex and underlying white matter) may impede the restoration of function in canonical language networks, and result in a less effective compensatory strategy involving the recruitment of non-canonical fronto-temporo-parietal networks in the right hemisphere.

Notably, our findings extend previous evidence suggesting that posterior temporal lesions might disrupt the restoration of function in canonical language networks (Heiss et al., 1999). In a longitudinal investigation comparing language recovery among groups of aphasic patients damage to frontal, sub-cortical, or temporal lesions during the first two months of recovery, Heiss and colleagues (1999) found that patients with temporal lesions showed limited improvement that was accompanied by both (1) a maintenance of right fronto-temporal language activation, and (2) a failure to re-activate left posterior temporal areas during language task performance (Heiss et al. 1999). Thus, Heiss and colleagues proposed that the preservation of temporal portions of the left hemispheric language network, and their eventual re-integration into the network may be a pre-requisite for the restoration of overall network function and successful language recovery (1999). Our data suggest that not all left temporal lesion patterns lead to chronic dependence on right fronto-temporo-parietal networks for language task performance, as the pICA analysis did not reveal a significant association between fMRI IC3 and lesion IC4, which featured contributions from voxels in the left MTG/STG (Figure 2). Indeed, Figure 2A shows that while lesion IC4 featured contributions from primarily cortical and superficial white matter voxels, lesion IC5 featured the strongest contributions from voxels in the deep white matter underlying the left posterior temporo-parietal cortex. While exploratory, our fiber tracking of the voxels showing the most dominant contributions to lesion IC4 and lesion IC5 suggests that lesion patterns involving regions with the most dominant contributions to lesion IC5 may have the capacity to simultaneously disrupt multiple pathways (including the IFOF, MdLF, and all three segments of the AF) that connect distal portions of the canonical semantic network, whereas lesion patterns involving regions with the most dominant contributions to lesion IC4 are likely to only affect the MdLF (Figure 4B). Thus, activation patterns resembling fMRI IC3 may reflect an inability to restore canonical network function following the disruption of multiple long-range connections by lesion patterns resembling lesion IC5, and this may in turn lead to a prolonged dependence on non-canonical regions in the right hemisphere to accomplish language tasks and less successful recovery.

It is important to note that an earlier PET study comparing activation evoked during speech by between patients with vs. without damage to the left IFGpop found that patients with damage to left IFGpop showed increased activation in the contralateral right IFGpop (Blank et al., 2003). Similar effects were reported by a recent meta-analysis by Turkeltaub and colleagues (2011). While loadings on the frontal lesion components (lesion IC2 and lesion IC3) did not showed significant correlations with loadings on fMRI IC3, loadings on lesion IC3 (which featured lesion contributions from the left IFGpop) showed a negative correlation with loadings on fMRI IC5 (which featured positive contributions from the right IFGpop) that did not survive multiple comparisons for the 25 tests performed ($r=0.37$, $p=0.01$ uncorrected; FWEp threshold = 0.002). Thus, while our data do not allow us to draw strong conclusions about the relationship between left IFGpop lesions and right frontal activation, they do provide evidence that is nonetheless consistent with the results of previous studies investigating this question.

*4.2 Limitations and future directions*

One limitation of this study is that it only included chronic patients. Thus, we cannot draw conclusions regarding whether activation patterns resembling fMRI IC3 reflect the long-term maintenance of early right hemispheric compensation into the chronic recovery phase, or if they reflect the development of a large-scale compensatory network over time. Further, while our data suggest that lesion patterns resembling lesion IC5 are associated with poorer performance on language tests in the chronic phase, it is unclear if this reflects severe initial impairment, impeded recovery, or both. Future studies using longitudinal designs are necessary to answer this question. We do note, however, that a recent study investigating how treatment-related improvements in naming abilities relate to lesion location found evidence that damage to the posterior temporal and inferior parietal deep white matter is associated with poor treatment response (Fridriksson 2010), suggesting that lesions affecting regions contributing to lesion IC5 may have a negative impact on recovery associated with therapy.

A second limitation is that the lesion ICs are relatively coarse in their spatial resolution, and the thresholded IC maps correspond to relatively broad swaths of tissue that would likely be affected by infarcts occurring in different branches of the left middle cerebral artery territory. Indeed, it has been previously noted that aphasia syndromes and their clinical characteristics are constrained by the vasculature of the LMCA territory (Rorden et al. 2007; Richardson et al. 2012; Henseler et al. 2014; Yourganov et al. 2015), and given that independent component analysis is a data-driven technique that decomposes MRI data into statistically independent spatial patterns, a high correspondence between the decomposed lesion data and the vasculature of the LMCA territory is to be expected. Nonetheless, the voxel contributions to each component are not uniform, indicating that specific cortical/sub-cortical areas contribute more to each component than to others.

Despite these limitations, data fusion techniques such as pICA have important applications in future aphasia research. For example, future studies might use such techniques investigate relationships between lesion patterns and other structural measures in stroke patients. This might shed light on whether increases in right hemispheric grey matter (Xing et al. 2015) and white matter (Pani et al. 2016) that have been recently reported in chronic stroke patients are related to the preservation of specific left hemispheric structures. While these questions are beyond the scope of the current study, they are questions that would be difficult to approach using standard mass univariate MRI analyses. Multivariate data fusion methods thus provide powerful tools for approaching such questions, and should be employed by future studies to enable a more comprehensive understanding of the structure-function relationships involved in the recovery of language after left hemispheric stroke.

## 5. Conclusions

Very little research has focused on how differences among stroke patients in the site of structural damage relate to differences in the functional networks recruited during the performance of language tasks. Here, we applied parallel ICA to the investigation of relationships between fMRI language task activation patterns and patterns of structural damage in patients with chronic post-stroke aphasia. Our results indicate that damage to the left posterior temporo-parietal cortex and underlying deep white matter is associated with reduced activity in bilateral regions that are typically activated during semantic processing, and with the recruitment of a non-canonical network that predominantly features right frontal, temporal, and parietal areas. Damage to the left posterior temporo-parietal cortex and underlying white matter might disable the restoration of canonical network function via the simultaneous disruption of structural pathways linking distal portions of canonical language networks.

**Acknowledgements**
Amber Martin
Christi Banks
NIH R01 HD068488
NIH R01 NS048281
Data were provided [in part] by the Human Connectome Project, WU-Minn Consortium (Principal Investigators: David Van Essen and Kamil Ugurbil; 1U54MH091657) funded by the 16 NIH Institutes and Centers that support the NIH Blueprint for Neuroscience Research; and by the McDonnell Center for Systems Neuroscience at Washington University.

# Supplementary Material

## Supplementary Table 1. Patient characteristics

| Patient | Age | Sex | EHI | TSS | BNT | COWAT | SFT | Average Fluency | %SD Correct |
|---|---|---|---|---|---|---|---|---|---|
| 1 | 63 | F | 0.55 | 1.0 | 59 | 27 | 47 | 37 | 68 |
| 2 | 78 | F | 1.00 | 4.1 | 57 | 21 | 39 | 30 | 50 |
| 3 | 41 | F | 0.50 | 5.8 | 9 | 4 | 9 | 6.5 | 38 |
| 4 | 54 | M | 1.00 | 1.6 | 7 | 5 | 6 | 5.5 | 66 |
| 5 | 46 | M | 0.90 | 1.0 | 53 | 20 | 42 | 31 | 46 |
| 6 | 52 | M | 0.58 | 1.0 | 60 | 11 | 27 | 19 | 48 |
| 7 | 56 | M | 1.00 | 3.4 | 32 | 2 | 14 | 8 | 50 |
| 8 | 53 | M | 1.00 | 5.0 | 50 | 8 | 20 | 14 | 76 |
| 9 | 55 | M | 1.00 | 1.2 | 58 | 15 | 30 | 22.5 | 72 |
| 10 | 48 | M | 1.00 | 6.1 | 22 | 0 | 12 | 6 | 20 |
| 11 | 63 | M | 1.00 | 1.0 | 60 | 8 | 10 | 9 | 40 |
| 12 | 56 | F | 1.00 | 1.0 | 33 | 6 | 20 | 13 | NA |
| 13 | 23 | M | 1.00 | 1.0 | 60 | 36 | 62 | 49 | 76 |
| 14 | 50 | M | 1.00 | 1.0 | 2 | 0 | 2 | 1 | 28 |
| 15 | 48 | F | 1.00 | 1.0 | 60 | 24 | 60 | 42 | 72 |
| 16 | 70 | F | 1.00 | 2.0 | 11 | 3 | 3 | 3 | 0 |
| 17 | 68 | M | 0.91 | 3.3 | 9 | 4 | 10 | 7 | 50 |
| 18 | 59 | M | 0.82 | 1.0 | 53 | 19 | 35 | 27 | 72 |
| 19 | 23 | F | 1.00 | 1.0 | 59 | 20 | 45 | 32.5 | 76 |
| 20 | 24 | F | 1.00 | 1.0 | 59 | 31 | 45 | 38 | 58 |
| 21 | 78 | F | 1.00 | 3.4 | 58 | 9 | 21 | 15 | 36 |
| 22 | 65 | M | 1.00 | 14.0 | 55 | 18 | 36 | 27 | 68 |
| 23 | 58 | F | 1.00 | 13.0 | 40 | 13 | 12 | 12.5 | 36 |
| 24 | 72 | F | 1.00 | 1.5 | 0 | 0 | 0 | 0 | NA |
| 25 | 50 | M | 1.00 | 2.9 | 0 | 0 | 0 | 0 | 42 |
| 26 | 57 | M | 1.00 | 2.1 | 2 | 2 | 1 | 1.5 | 48 |
| 27 | 51 | M | 1.00 | 1.1 | 37 | 8 | 15 | 11.5 | 66 |
| 28 | 43 | M | 1.00 | 1.3 | 50 | 11 | 22 | 16.5 | 44 |
| 29 | 24 | M | 0.83 | 2.3 | 21 | 9 | 14 | 11.5 | 2 |
| 30 | 67 | F | 1.00 | 2.2 | 6 | 2 | 3 | 2.5 | 50 |
| 31 | 62 | F | -1.00 | 4.4 | 33 | 20 | 23 | 21.5 | 62 |
| 32 | 44 | F | 0.91 | 2.1 | 41 | 7 | 18 | 12.5 | 30 |
| 33 | 62 | M | 1.00 | 2.6 | 54 | 10 | 19 | 14.5 | 63 |
| 34 | 31 | M | 1.00 | 4.8 | 21 | 7 | 7 | 7 | NA |
| 35 | 61 | M | 1.00 | 9.6 | 25 | 1 | 6 | 3.5 | NA |
| 36 | 64 | M | -1.00 | 2.7 | 51 | 1 | 16 | 8.5 | 30 |
| 37 | 38 | F | 0.91 | 1.8 | 46 | 11 | 12 | 11.5 | 72 |
| 38 | 53 | F | 1.00 | 9.2 | 34 | 2 | 7 | 4.5 | 74 |
| 39 | 54 | M | 0.92 | 3.3 | 33 | 5 | 19 | 12 | 55 |

| 40 | 46 | M | 1.00 | 1.3 | 31 | 1 | 9 | 5 | 34 |
| 41 | 90 | F | 0.71 | 1.3 | 3 | 0 | 0 | 0 | 0 |
| 42 | 29 | F | 1.00 | 3.4 | 50 | 4 | 19 | 11.5 | 0 |
| 43 | 67 | M | 1.00 | 12.4 | 34 | 2 | 16 | 9 | 26 |

*EHI – Edinburgh handedness inventory, TSS – time since stroke, BNT – Boston naming test, COWAT – controlled oral word association test, SFT – semantic fluency test, %SD Correct -- % Semantic Decision Correct.
*Note:* The average fluency score is the average of the COWAT and SFT.

## Supplementary Analyses 1-3

**Rationale:** Joint ICA (jICA) is a statistical method that is often used for identifying structure-function relationships that differ between patient and control groups (Calhoun et al., 2006). JICA component selection is typically based on the results of statistical tests for between-group differences in mixing coefficients for each joint component (Calhoun, Adali, Giuliani, et al. 2006; Specht et al. 2009; Abel et al. 2015). This approach is reasonable when the structural differences between groups are subtle and/or of theoretical interest (i.e. for differentiating patients with schizophrenia from healthy individuals – Calhoun et al., 2006). However, when comparing healthy controls to groups such as chronic stroke patients, the differences in brain structure (within the group-level lesion territory) between patients and controls are not typically of interest because they are relatively obvious (i.e. chronic stroke lesions can be identified visually), and not directly useful for drawing conclusions about lesion-behavior relationships (notably, lesion-behavior mapping methods focus on the variation in damage within groups of patients). Because jICA component selection is based on identifying components with mixing coefficients that differ most substantially between patient and control groups, we hypothesized that jICA component selection would be biased towards selecting joint components with lesion features containing high voxel loadings on voxels that are most frequently lesioned in the patient group. If this were the case, then the lesion feature of the maximally differentiating joint IC selected following jICA should be a close transformation of lesion frequency information. This information is relevant because previous studies (Specht et al., 2009; Abel et al., 2015) have utilized the jICA approach to assess lesion-activation relationships in chronic stroke patients. We test these predictions in S1.

In addition, previous studies (Specht et al., 2009; Abel et al., 2015) have used cerebrospinal fluid (CSF) tissue probabilistic maps (TPMs) as indicators of lesion status for jICA analyses. Previous work by our lab (Griffis et al. 2016) suggests that CSF TPM estimates alone are not optimal predictors of voxel lesion statuses. This could result in incorrect spatial inferences when interpreting component features that are based on CSF values. Thus, we aimed to assess the spatial similarity of masks based on CSF TPM estimates to ground truth lesion masks, and to compare CSF spatial similarity results to those obtained when using TPMs that were estimated based on ground truth spatial lesion priors. We also assessed whether relationships identified with pICA would be of similar strength when using CSF TPMs as lesion indicators. These assessments are presented in S2.

To follow up on the results from S1 and S2, we performed jICA using CSF TPMs as lesion indicators in order to assess whether component selection is biased by voxel lesion frequencies when CSF TPMs are used instead of lesion masks. Finally, based on the results of S2, we assessed whether jICA component selection is biased towards components with high CSF feature loadings at voxels that have high CSF values at the group level. These analyses are presented in S3.

A detailed patient-level characterization is provided in S4.

**S1. Lesion frequency effects in Joint ICA (jICA) of lesion and fMRI data**

**Participants:** Patients used in the main analysis were utilized in this analysis along with 43 healthy controls selected to minimize differences in demographic variables (see Supplementary Table 2). fMRI data were acquired using the same SDTD task and processed using the same pipeline described for the patients in the main text.

**Supplementary Table 2. Group demographics**

| Group | N | Age | Sex | Handedness |
|---|---|---|---|---|
| Patients | 43 | 53 (15) | 25 M | 0.85 (0.43) |
| Controls | 43 | 54( 14) | 23 M | 0.80 (0.41) |

**jICA with lesion data:** Joint ICA was performed using the Fusion ICA toolbox for SPM. Features were identical to those used in the pICA analysis described in the main text (lesion masks and contrast estimate maps). Lesion mask inputs for the control participants consisted of nifti volumes containing only zero valued voxels. The minimum description length algorithm was used to estimate the dimensionality of the stacked data (Calhoun, Adali, Kiehl, et al. 2006), and the number of components to be derived was set to 6. Joint ICA was performed using default settings. A two-sample t-test was used to select joint components that differed significantly in mixing coefficients between the groups. The IC with the most reliable group difference in loading coefficients was joint IC6 ($t_{84}$=9.37, p<0.001). The lesion feature for this component was then thresholded at |z|>1.9 (i.e. identical threshold to that used in the primary analyses) for visualization purposes, and is shown along with the group lesion-frequency map (thresholded at 75% max frequency = 24) in Supplementary Figure 1A. Spearman rank correlation was used to assess whether voxel loadings for the lesion feature of joint IC6 were a monotonic function of voxel lesion frequencies. This revealed a strong monotonic relationship between voxel loadings for the lesion feature of joint IC6 and voxel lesion frequencies (rho = 0.96, p<0.001). A plot illustrating the nature of this relationship is shown in Supplementary Figure 1B. We also assessed the relationship between voxel loadings for the lesion IC identified by the pICA analysis (lesion IC5) and voxel lesion frequencies using an identical procedure. This revealed a weak monotonic relationship between voxel loadings for lesion IC5 and voxel lesion frequencies (rho = 0.20, p<0.001). A scatterplot illustrating the nature of this relationship is shown in Supplementary Figure 1C. Thus, our supplementary analysis suggests that joint component selection for the jICA method is biased by voxel lesion frequencies when comparing stroke patients to healthy controls. In addition, it suggests that voxel lesion frequencies do not strongly influence component selection for the pICA method. Finally, rank correlation was used to assess whether voxel loadings for the fMRI feature of joint IC6 (Supplementary Figure 1D, right panel) were a monotonic function of the group-level voxel estimates for the SD-TD GLM contrast

(Supplementary Figure 1D, left panel), revealing a moderate relationship (rho = 0.46, p<0.001; Supplementary Figure 1D, bottom panel).

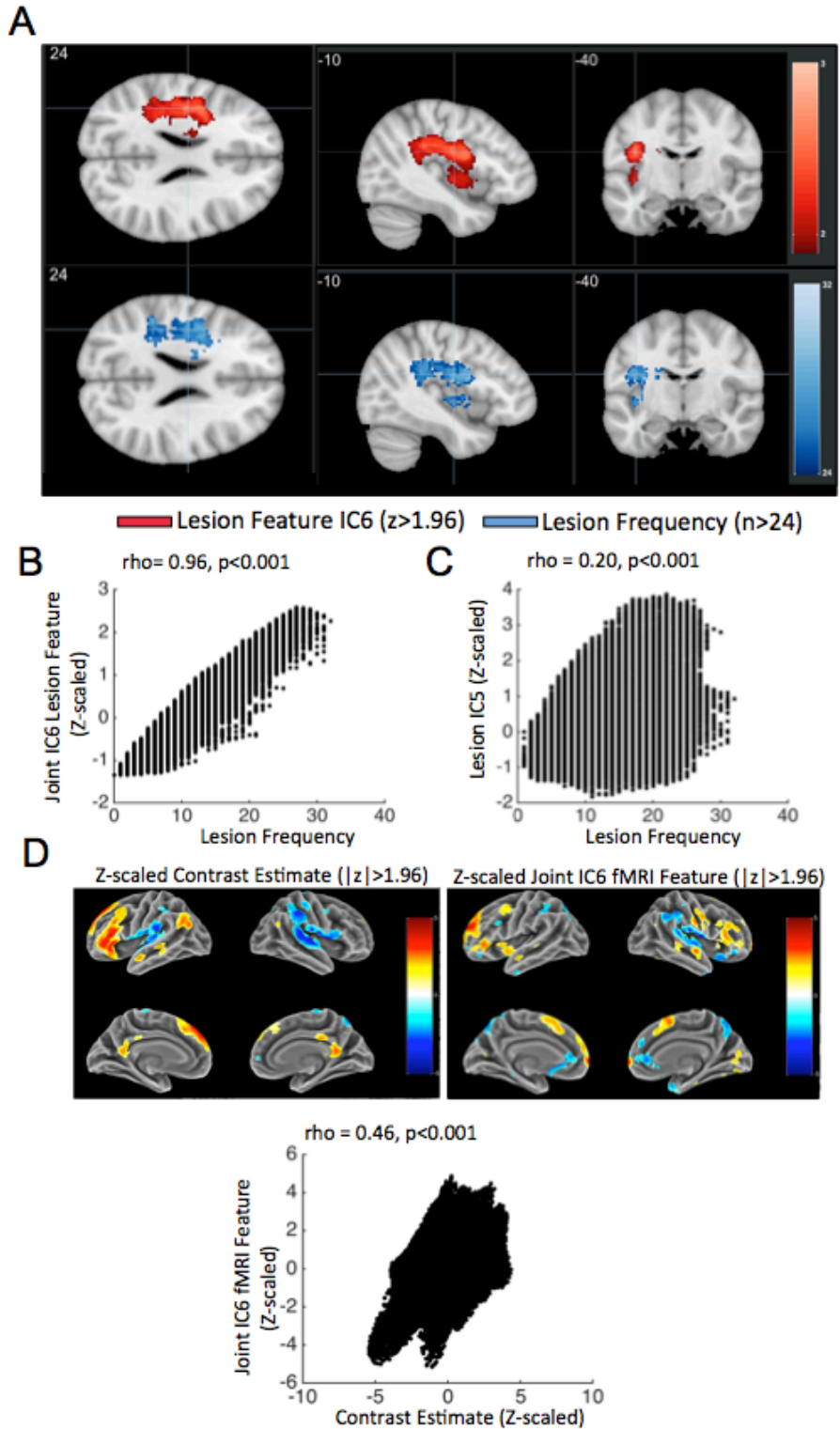

**Supplementary Figure 1. A.** The thresholded (|z|>1.9) spatial map for the lesion feature of joint IC6 (i.e. the component with the largest between-group difference in loading coefficients) is shown on the top in red, and the thresholded (n>24 – 75% max frequency)

spatial map for lesion frequencies across all 43 patients is shown on the bottom in blue. Note that the images shown are nearly identical. **B.** A scatterplot illustrating the relationship between the number of patients lesioned at a given voxel (x-axis) and the z-scaled loading for the same voxel in the lesion feature of joint IC6 obtained from the jICA analysis. Rank correlation analysis revealed that voxel loadings and lesion frequencies were strongly correlated (rho = 0.96). **C.** A scatterplot illustrating the relationship between the number of patients lesioned at a given voxel (x-axis) and the z-scaled loading for the same voxel in lesion IC5 obtained from the pICA analysis. Rank correlation analysis revealed that voxel loadings and lesion frequencies were only weakly correlated (rho = 0.20). **D.** Spatial maps for the z-scaled contrast estimate map for the SD-TD GLM contrast from the stroke patients (left) and the z-scaled joint IC6 fMRI feature (right). Both maps are thresholded at $|z| > 1.96$. The scatterplot below illustrates the relationship between the z-scaled contrast estimate map and the z-scaled joint IC6 fMRI feature. Rank correlation analyses revealed that the fMRI features were moderately correlated (rho=0.46).

**Conclusion:** Component selection for jICA of lesion and fMRI data obtained from groups of stroke patients and healthy controls is likely to be biased to select for components with lesion features that contain high loadings on voxels with high lesion frequencies in the patient group. Component selection in pICA does not appear to be affected by voxel lesion frequencies.

**S2. Comparison of CSF and lesion tissue probabilistic map (TPM) spatial similarities to lesion masks**

**Aims:** To assess the spatial similarity of information provided by CSF TPMs to ground truth lesion masks, and compare it with the spatial similarity of information provided by TPMs that were defined based on ground truth lesion information. To assess pICA performance when using CSF TPMs as lesion indicators.

**Participants:** Patients used in the main analysis were utilized in this analysis.

**CSF TPMs:** CSF TPMs were obtained using the unified segmentation and normalization routine in SPM12 with default settings. Smoothed CSF TPMs were created by smoothing the CSF TPMs with an 8mm full-width half maximum (FWHM) Gaussian kernel.

**Lesion TPMs:** Lesion TPMs were obtained using the unified segmentation and normalization routine in SPM12 with an additional lesion tissue class. Spatial priors for the lesion tissue class were defined using each patient's posterior probability map from automated lesion classification (similar to the procedure used by (Sanjuan et al. 2013)). Smoothed lesion TPMs were created by smoothing the CSF TPMs with an 8mm full-width half maximum (FWHM) Gaussian kernel.

**Spatial similarity analysis:** Smoothed and unsmoothed CSF/Lesion TPMs were thresholded at 25%, 50%, and 75% tissue probability thresholds and binarized. To assess the spatial similarity between the binarized TPMs and the ground truth lesion masks,

Dice similarity indices (DSIs) were computed between the binarized TPMs from each patient and the patient's binarized lesion mask. The DSI for each binarized TPM is defined as: $DSI_{TPM,Mask} = (2 \cdot (TPM \cap Mask))/(TPM+Mask)$. Higher DSIs indicate higher spatial similarity. Wilcoxon signed rank tests were used to assess differences in DSIs between CSF and Lesion TPMs. Boxplots illustrating results are shown in Supplementary Figure 2.

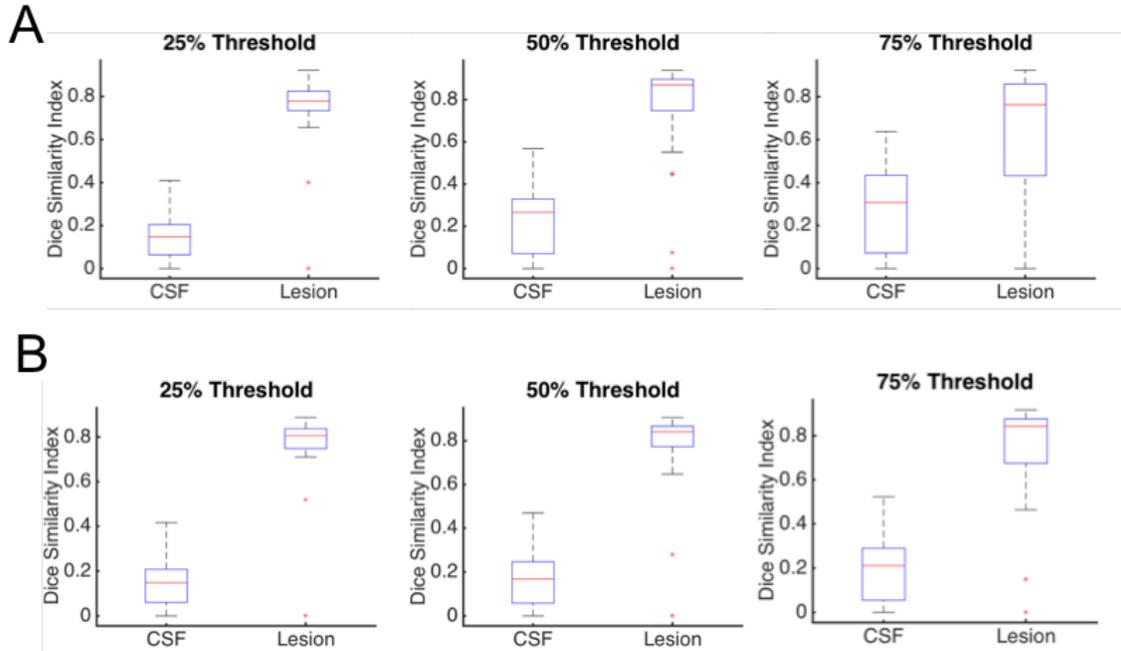

**Supplementary Figure 2.** Spatial similarity analysis results. Median DSIs were higher for lesion TPMs than CSF TPMs at all three thresholds with (A) and without (B) spatial smoothing (all $p<0.001$). Median DSIs for CSF TPMs were below 0.6 (typically considered "good" spatial similarity – e.g. Seghier et al., 2008) for all thresholds with (A) and without (B) smoothing.

**Parallel ICA Analysis:** To determine if the low spatial similarity of CSF TPMs to lesion masks might influence the results of a pICA analysis using CSF TPMs as lesion indicators, pICA was performed using the smoothed (8mm FWHM) CSF TPMs as inputs for the lesion feature. The same lesion zone mask used for the lesion feature in the pICA analyses reported in the main text was used as the mask for the CSF lesion feature here. Analyses were performed using both (1) the same number of lesion ICs (5 ICs) used for the pICA in the main text, and (2) the number of ICs suggested by the application of the MDL algorithm to the CSF data (6 ICs). Component selection for each analyses utilized the same criteria (FWE-correction across all pairwise correlations between component loading coefficients) as reported in the main text. For both the pICA with 5 CSF ICs and the pICA with 6 CSF ICs, no significant correlations between loading coefficients on the CSF ICs and loading coefficients on the fMRI ICs were observed (max |r| = -0.34, FWEp = 0.65, and max |r| = 0.36, FWEp = 0.51, respectively).

**Conclusion:** CSF TPMs, both smoothed and unsmoothed, are sub-optimal sources of lesion information as indicated by the observation that median DSIs for CSF TPMs were generally low (all <0.6) and by the finding that median DSIs for CSF TPMs were significantly lower than those obtained for the lesion TPMs at all thresholds. Accordingly, structure-function relationships observed when performing pICA with CSF TPMs and fMRI contrast estimate maps as features did not reach even trend-level statistical significance.

**S3: Lesion frequency effects in Joint ICA (jICA) of CSF and fMRI data**

Because CSF TPMs were found to provide less reliable lesion status information than lesion TPMs, we next assessed whether lesion frequency biasses might be present when using CSF tissue probability maps (TPMs) as proxies for lesion information in jICA. Thus, a second joint ICA was performed using the Fusion ICA toolbox for SPM. Features consisted of CSF TPMs and fMRI contrast estimate (i.e. the same features used by Specht et al., 2009 and Abel et al., 2015). The minimum description length algorithm was used to estimate the dimensionality of the stacked data (Calhoun, Adali, Kiehl, et al. 2006), and the number of components to be derived was set to 8. Joint ICA was performed using default settings. CSF features were masked to only include voxels that were lesioned in at least one patient. A two-sample t-test was used to select joint components that differed significantly in mixing coefficients between the groups. The IC with the most reliable group difference in loading coefficients was joint IC6 ($t_{84}$=8.01 p<0.001). Spearman rank correlation was used to assess whether voxel loadings for the CSF feature of joint IC8 were a monotonic function of voxel lesion frequencies. This revealed a weak monotonic relationship between voxel loadings for the CSF feature of joint IC8 and voxel lesion frequencies (rho = 0.21, p<0.001). A plot illustrating the nature of this relationship is shown in Supplementary Figure 3A. Because in S2, we found evidence that CSF tissue probabilities are relatively poor indicators of lesion status, we hypothesized that jICA of CSF TPMs might instead be biased to select voxels with high CSF TPM probabilities across patients. Thus, we performed a one-sample t-test on the masked CSF TPMs from the stroke group to obtain the group-level contrast estimate map. We then used Spearman rank correlation to assess whether voxel loadings for the CSF feature of joint IC8 were a monotonic function of CSF contrast estimate values. This revealed a strong monotonic relationship between voxel loadings for the CSF feature of joint IC8 and CSF contrast estimates (rho=0.99, p<0.001). A plot illustrating this relationships is shown in Supplementary Figure 3B. Spatial maps corresponding to the CSF feature of joint IC8 (thresholded at |z|>1.9) and z-scaled CSF contrast estimates (thresholded at z>1.96) are shown in Supplementary Figure 3A. Finally, rank correlation was used to assess whether voxel loadings for the fMRI feature of joint IC8 (Supplementary Figure 3D, right panel) were a monotonic function of the group-level voxel estimates for the SD-TD GLM contrast (Supplementary Figure 3D, left panel), as performed for the jICA analysis in Supplementary Analysis 1. This revealed a moderately strong relationship (rho = 0.64, p<0.001; Supplementary Figure 3D, bottom panel).

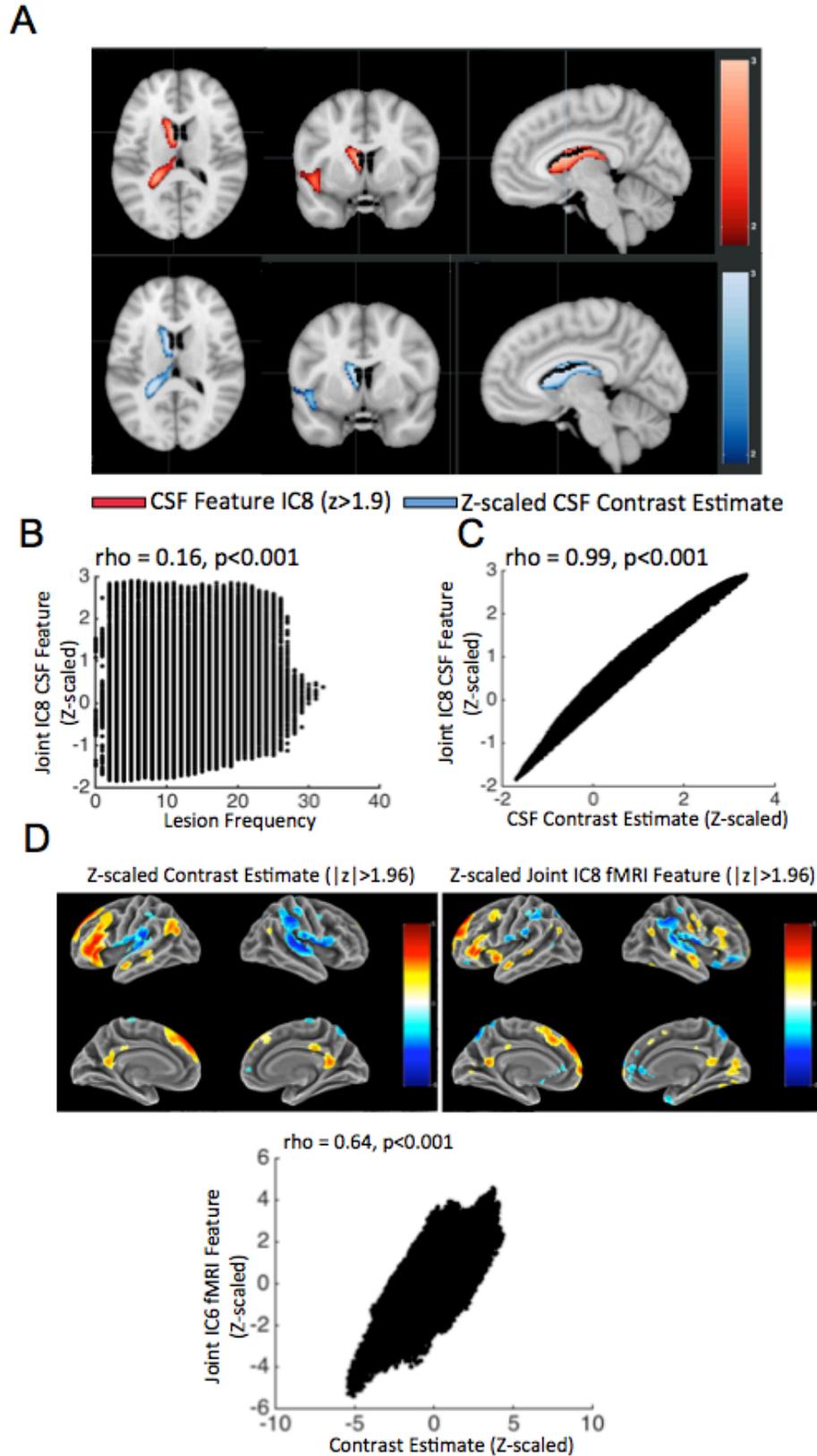

**Supplementary Figure 3. A.** The thresholded (|z|>1.9) spatial map for the CSF feature of joint IC8 (i.e. the component with the largest between-group difference in mixing

coefficients) is shown on the top in red, and the thresholded z-scaled CSF contrast estimate map is shown on the bottom in blue. Note that the images shown are nearly identical. **B.** A scatterplot illustrating the relationship between the number of patients lesioned at a given voxel (x-axis) and the z-scaled loading for the same voxel in the CSF feature of joint IC8 obtained from the jICA analysis. Rank correlation analysis revealed that voxel loadings and lesion frequencies were weakly correlated (rho = 0.21). **C.** A scatterplot illustrating the relationship between the z-scaled CSF contrast estimates at a given voxel (x-axis) and the z-scaled loading for the same voxel in the CSF feature of joint IC8 obtained from the jICA analysis. Rank correlation analysis revealed that voxel loadings summed CSF TPM estimates were nearly perfectly correlated (rho = 0.99). **D.** Spatial maps for the z-scaled contrast estimate map for the SD-TD GLM contrast from the stroke patients (left) and the z-scaled joint IC8 fMRI feature (right). Both maps are thresholded at |z| > 1.96. The scatterplot below illustrates the relationship between the z-scaled contrast estimate map and the z-scaled joint IC8 fMRI feature. Rank correlation analyses revealed that the fMRI features were strongly correlated (rho=0.64).

**Conclusion:** Component selection for jICA of CSF and fMRI data obtained from groups of stroke patients and healthy controls is likely to be biased to select for components with CSF features that contain high loadings on voxels with high CSF estimates in the patient group. Consistent with the finding from S2 that CSF TPM estimates are poor indicators of voxel lesion status, aggregate CSF TPM estimates are only weakly correlated with voxel lesion frequencies. However, consistent with the finding from S1 that jICA component selection is biased by voxel lesion frequencies in stroke patients, jICA component selection is biased to select for components with lesion features that have high loadings from voxels with consistently high CSF TPM estimates across patients.